\RequirePackage{silence}
\documentclass[paper]{JFM-FLM_Au}

\usepackage{silence}
\usepackage{comment}
\usepackage{siunitx}
\usepackage{graphicx}
\usepackage{float}
\usepackage{placeins}

\AddToHook{env/appen/begin}{}

\makeatletter
\def\jfm@pageno{{\cppagefont\thepage}}
\def\oddabsfooterflag{\hbox to\textwidth{\hfill\jfm@pageno}}
\def\@oddfoot{\oddabsfooterflag}
\def\evenabsfooterflag{\hbox to\textwidth{\jfm@pageno\hfill}}
\def\pagelimitfooter{\hbox to\textwidth{\jfm@pageno\hfill}}
\makeatother
\definecolor{blue}{rgb}{0.00,0.00,1.00}
\definecolor{orange}{rgb}{0.00,0.00,0.00}

\lefttitle{S. V. Kethanur Balasubramaniam, N. N. Chamala and O. Coutier-Delgosha}
\righttitle{Journal of Fluid Mechanics}

\title{Cavitation inception delayed by structured surface roughness}

\author{Sanjay Vasanth Kethanur Balasubramaniam\aff{1}, Naga Nitish Chamala\aff{1} \and Olivier Coutier-Delgosha\aff{1}}

\affiliation{\aff{1}Kevin T. Crofton Department of Aerospace and Ocean Engineering, Virginia Tech, Blacksburg, USA}

\corresau{Sanjay Vasanth Kethanur Balasubramaniam, kbsanjayvasanth@vt.edu, Olivier Coutier-Delgosha, ocoutier@vt.edu}

\begin{document}
\maketitle

\begin{abstract}
Cavitation inception over structured rough surfaces is investigated experimentally in the diverging section of a Venturi. The diverging wall carries laser-etched checkerboards of square cavities with roughness aspect ratios $0.03 \leq k/d \leq 0.72$, tested at a Reynolds number of 95\,160. Inception cavitation numbers were measured together with high-speed shadowgraphs of the near-wall flow, from which upstream-migrating nuclei were tracked and vapor-masked velocity fields were obtained by optical flow. Smooth walls recessed by the etched volume served as a control. The roughness delays inception, most strongly at intermediate $k/d$, where the inception cavitation number falls from 4.07 on the smooth wall to 3.33--3.46, whereas the recessed walls show no delay. The supply of nuclei is similar on all surfaces, so the delay does not come from a lack of nuclei. Instead, the roughness changes the near-wall flow. For $0.08 \leq k/d \leq 0.444$, a mean reverse-flow layer forms along the wall beneath a lifted shear layer, and vapor at inception appears further from the throat and away from the wall. At $k/d = 0.72$ the flow reverses only intermittently and the delay is smaller. On all surfaces except $k/d = 0.08$, the local cavitation number at inception varies linearly with the near-wall fluctuation energy, consistent with nuclei activated by pressure excursions about 6--8 times the r.m.s.\ fluctuation. The roughness thus delays inception not by limiting the nuclei, but by weakening the near-wall fluctuations that reach them.
\end{abstract}


\section{Introduction}
\label{sec:intro}

Cavitation inception, the first appearance of vapor in an initially liquid flow, requires two ingredients: a nucleus, typically a microscopic gas bubble or a gas pocket trapped on a surface, and low-pressure events that are deep enough and last long enough to make it grow explosively \citep{Franc2004}. Inception often sets the operating limits of pumps, turbines, propellers and hydraulic structures, in which cavitation causes noise, vibration, surface erosion and loss of efficiency \citep{Arndt1981}. In turbulent flows the mean pressure at inception usually remains above the vapor pressure, and growth is triggered instead by transient low-pressure events, most often in the cores of stretched vortices \citep{Katz1986,Belahadji1995,Ram_Agarwal_Katz_2020,Agarwal2023}. Inception therefore depends as much on the nuclei as on the flow. The effect of water quality on inception has long been recognized \citep{Arndt1992}, and experiments with controlled nuclei populations have shown that inception depends strongly on the nuclei available \citep{Brandner2022,Russell2023}, with the rate of inception events proportional to the rate at which nuclei are supplied \citep{Allan2023,Allan2025}. \citet{Allan2025} described inception as the confluence of two unlikely events: a sufficiently strong low-pressure event (in their flow, a vortex core below the vapor pressure) and a sufficiently weak nucleus at the same place and time.

Near walls, the supply and fate of nuclei are tied closely to the boundary layer. In flows that develop attached cavitation, such as over curved walls and hydrofoils, \citet{Ram_Agarwal_Katz_2020} showed that inception is sustained by a feedback loop: residual microbubbles left by collapsing cavities migrate upstream within the low-momentum near-wall region, grow by gas diffusion and seed new cavitation events. \citet{Li2026}, seeding single nuclei on demand in a Venturi, found that attached cavitation forms when a bubble enters the part of the boundary layer in which the pressure is below the vapor pressure. In both cases, the position of the nuclei relative to the near-wall flow decides whether, and where, they cavitate.

The walls of hydraulic machines are rarely hydraulically smooth \citep{Arndt1981}. Wall roughness alters precisely this near-wall flow \citep{Jimenez2004,Kadivar2021}, and its effect on cavitation is not simple. In an attached turbulent boundary layer over regular triangular grooves, \citet{ArndtIppen1968} found that roughness makes cavitation appear earlier: the inception index scaled with the skin-friction coefficient, $\sigma_i \approx 16\,C_f$, because roughness raises the wall shear and with it the pressure fluctuations. Roughness can also delay inception. Applied to the tip region of a hydrofoil, it delays tip-vortex cavitation by weakening the vortex \citep{Asnaghi2019,Svennberg2020}, and a similar delay has been predicted for tip-leakage flows \citep{Feng2026}. For developed cavitation, roughness and surface texturing have been reported to modify the dynamics of sheet cavities in Venturis \citep{CoutierDelgosha2005,Danlos2014} and to suppress cavitation around a cylinder \citep{Nichik2024}. Roughness can also act on the nuclei themselves: gas trapped in surface crevices can serve as nuclei, with a stability that depends on the crevice geometry and on wettability \citep{Atchley1989,Borkent2009}. A rough wall may therefore change both ingredients of inception at once, the low-pressure events near the wall and the nuclei available to them. How these effects combine, and which of them dominates, remains unknown, because the near-wall flow and the behavior of individual nuclei have not been measured together over rough walls.

This paper addresses this question in the strongly decelerating near-wall flow in the diverging section of a Venturi, whose diverging wall carries laser-etched surfaces with a regular, periodic array of cavities and a roughness aspect ratio $0.03 \leq k/d \leq 0.72$. We refer to these as structured rough surfaces, to distinguish them from random, sand-grain-type roughness. On these surfaces, inception requires a lower cavitation number than on a smooth wall, especially at intermediate roughness. To find why the structured roughness delays inception, we combine inception measurements at a fixed flow rate, a recessed-wall control study, the tracking of upstream-migrating nuclei and vapor-masked, time-resolved velocity fields from optical flow \citep{Teed2020}. The tracking and the velocity fields address the two ingredients of inception identified above. The tracking tests whether the roughness changes the supply of nuclei, and the velocity fields test whether it changes the near-wall flow that produces the low-pressure events reaching them. Following \citet{ArndtIppen1968}, we then relate the local cavitation number at inception to the near-wall fluctuation energy, to test whether a single fluctuation-based criterion accounts for the inception condition on all surfaces.

\section{Experimental set-up}\label{sec:setup}

\subsection{Facility and test section}\label{sec:facility}

\begin{figure}
  \centerline{\includegraphics[width=0.9\textwidth]{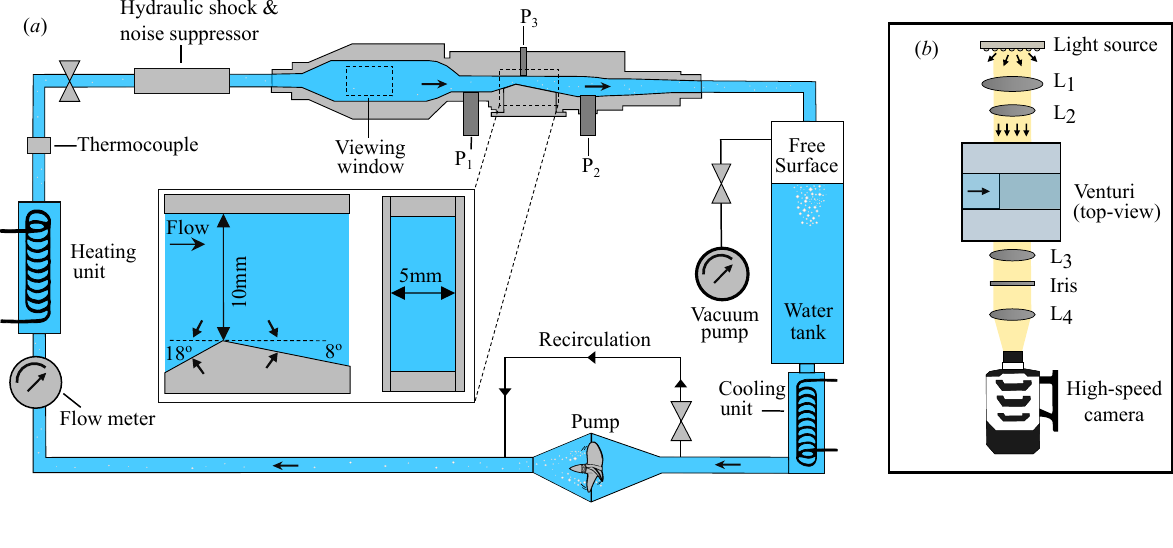}}
  \caption{Experimental set-up. (\textit{a}) Schematic of the cavitation loop and test section. The loop pressure is lowered with the vacuum pump on the storage tank, and the static pressures are measured at $P_1$ and $P_2$ (the tap $P_3$ is not used in this study). Inset: side view of the Venturi, with the $18^{\circ}$ converging and $8^{\circ}$ diverging sections and the 10~mm throat, and its 5~mm wide cross-section. (\textit{b}) Optical arrangement for the side-view imaging, seen from above. Light from the LED source passes through a condenser lens $L_1$ and an achromat $L_2$, crosses the Venturi along its span, and is imaged onto the high-speed camera by the achromats $L_3$ and $L_4$, with an iris between them.}
\label{fig:expsetup}
\end{figure}

The experiments were performed in a closed-loop cavitation mini-channel with a Venturi test section (figure~\ref{fig:expsetup}\textit{a}), an upgraded version of the design of~\citet{CoutierDelgosha2009}. The test section, which holds the Venturi inserts and the pressure ports, was printed in Digital ABS with a polyjet printer (Stratasys Objet500 Connex-3), whose 14~$\mu$m build layers make it watertight at high flow rates. The pump was driven by a variable-frequency drive, and a secondary recirculation loop kept its operation stable at low flow rates. Before testing, the storage tank was filled and left undisturbed for 72~h so that large bubbles could rise and escape at its free surface, which was maintained during operation. Naturally occurring micronuclei remained in the water. An external cooling loop kept the water temperature constant.

The Venturi inserts were printed by the same process in VeroUltraClear for optical access. The channel is 5~mm wide. Its height contracts from 20~mm at the inlet to $H = 10$~mm at the throat (area ratio 2) through an $18^{\circ}$ converging section and then expands through an $8^{\circ}$ diverging section. The diverging section continues beyond the insert into the test section, and only its upstream part, on the insert, carries the surfaces studied here. Removable acrylic side windows provide optical access and set the channel width. They also allow the Venturi insert to be exchanged. Each surface was made on its own insert, and the inserts were swapped between tests while the rest of the test section, the water and the operating procedure stayed the same.

\subsection{Structured rough surfaces}\label{sec:surfaces}

\begin{figure}
  \centerline{\includegraphics[width=0.9\textwidth]{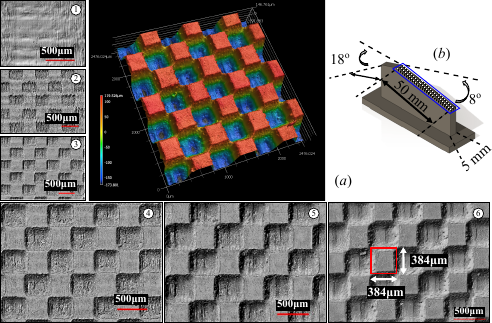}}
  \caption{Structured rough surfaces. (\textit{a}) Three-dimensional height map of Rough~5, measured over the etched area near the throat with the laser-scanning confocal microscope: the unetched cells (red) are the original wall and the etched cells (blue) lie below it. (\textit{b}) Venturi insert, showing the etched region (blue outline). (1)--(6) Laser confocal micrographs of the smooth surface (1) and of Rough~1--5 (2--6), showing the checkerboard of etched cells of side $d = 384~\mu$m.}
\label{fig:roughness}
\end{figure}

The diverging wall of the Venturi was roughened by laser etching with an Epilog Zing 16/24 laser engraver, so that only the surface changed while the bulk geometry stayed the same. The etching was applied in a checkerboard pattern of square cells of side $d = 384~\mu$m, so that adjacent etched cells are separated by an unetched gap of width $d$ (figure~\ref{fig:roughness}). The checkerboard gives a periodic, reproducible texture with a known length scale. Its cells are separate, so there are no continuous grooves that could guide the near-wall flow in one direction. Jarvis error-diffusion dithering was applied at an etching resolution of 1000~dpi and a pulse frequency of 2500~Hz. This distributed the laser pulses irregularly over the etched cells, giving a statistically homogeneous texture without banding along the raster path.

The laser power was held at 10\,\% of its maximum and the scan speed was varied from 5\,\% to 100\,\% of its maximum, so that the only parameter changed was the energy deposited per unit length of scan, which sets the depth of ablation: slower scans deposit more energy and produce deeper, rougher etching. Calibration surfaces etched over this range were measured with a laser-scanning confocal microscope (Keyence VK-X3000) in terms of $Ra$, $Rz$ and $Sa$, with etched volumes obtained using the MultiFileAnalyzer software. From these tests, a calibration curve relating roughness to scan speed was obtained and used to select the scan speed of each test surface. Six surfaces were tested: a smooth reference, obtained by sanding the raw printed surface with fine-grit abrasive paper, and five structured rough surfaces. Rough~1, etched at 100\,\% scan speed, is the lightest and Rough~5, at 20\,\%, the deepest. The roughness is characterized by the aspect ratio $k/d$, with $k = Rz$, a purely geometric ratio of the etched cells that ranges from 0.03 on the smooth surface to 0.72 (table~\ref{tab:roughness_surfaces}).

An inner-scaled roughness height $k_s^+$ is not used, since it requires the friction velocity of an attached boundary layer, whereas the etched wall lies entirely in the decelerating flow of the diverging section, where the wall shear stress is small or, on Rough~1--4, reversed (\S\ref{sec:flow}). The roughness is also small compared with the channel, with $k/H = 0.001$--$0.028$.

\begin{table}
    \centering

    \begin{tabular}{lccccc}
        \hline
        Surface & Laser speed (\%) & $Sa$ ($\mu$m) 
        & $Rz$ ($\mu$m) & $k/d$ & $\theta$ ($^{\circ}$) \\
        \hline
        Smooth   & ---   & $2.11$  & $11.64$  & $0.030$ & $80 \pm 1$ \\
        Rough 1  & $100$ & $5.11$  & $31.02$  & $0.080$ & $79 \pm 3$ \\
        Rough 2  & $70$  & $12.06$ & $54.32$  & $0.141$ & $77 \pm 5$ \\
        Rough 3  & $50$  & $25.44$ & $102.82$ & $0.267$ & $69 \pm 7$ \\
        Rough 4  & $30$  & $53.28$ & $170.50$ & $0.444$ & $66 \pm 6$ \\
        Rough 5  & $20$  & $90.14$ & $277.69$ & $0.720$ & $78 \pm 1$ \\
        \hline
    \end{tabular}

    \caption{Laser scan speed (percentage of maximum, at 10\,\% power) and roughness of the six surfaces. $Sa$ and $Rz$ are means of 20 measurements on the diverging wall, and $k/d = Rz/d$, with $d = 384~\mu$m the side of the checkerboard cells. $\theta$ is the apparent static contact angle, the mean of three sessile drops $\pm$ one standard deviation.}
    
    \label{tab:roughness_surfaces}
\end{table}

The wettability of each surface was characterized by its apparent contact angle $\theta$, measured with an optical tensiometer (Theta Flow, Biolin Scientific). Three 5~$\mu$L sessile drops of deionized water were placed at different positions, and $\theta$ was taken 30~s after deposition as the mean of the left and right angles (table~\ref{tab:roughness_surfaces}). All surfaces are wetting, with mean angles between $66^{\circ}$ and $80^{\circ}$. The angle does not vary monotonically with $k/d$, and the differences between surfaces lie within the drop-to-drop scatter, except possibly on Rough~4. Its role in inception is examined in \S\ref{sec:reversal}.


\subsection{Operating conditions and cavitation number}\label{sec:operating}

In initial runs, inception was reached on each surface by raising the flow rate with the tank at atmospheric pressure. It occurred at higher flow rates as the roughness increased, from $47.5~\mathrm{L\,min^{-1}}$ on the smooth surface to $54.5~\mathrm{L\,min^{-1}}$ on Rough~4, with Rough~5 incepting at a lower flow rate than Rough~4. These runs showed that the roughness delays inception. Each surface, however, incepted at a different flow rate, and hence at a different Reynolds number ($\Rey \approx 113\,000$--$130\,000$), so the effect of the roughness could not be separated from that of the flow. To quantify the delay under identical flow conditions, the main experiments were made at a fixed flow rate of $40~\mathrm{L\,min^{-1}}$. The pressure in the sealed loop was then lowered with a vacuum pump while the throat region was watched on the live high-speed video. The pump was stopped once cavitation events, after first appearing, recurred continuously instead of dying out, and the recordings were made at that condition. This visual criterion follows common practice in inception studies \citep{ArndtIppen1968}, and the transition is sharp, since cavitation becomes self-sustaining once the first events occur \citep{Allan2023}. The same criterion was applied to all surfaces by the same observer. The increasing-flow-rate protocol was used only for the control experiment of \S\ref{sec:delay}.

The flow rate was measured with a B\"urkert Type~SE32 turbine flow meter, with an accuracy of $\pm 0.15~\mathrm{L\,min^{-1}}$. Its reading fluctuated within $\pm 0.5~\mathrm{L\,min^{-1}}$, so this range is taken as the flow-rate uncertainty. The water temperature was kept at $23\,^{\circ}\mathrm{C}$ and monitored with a K-type thermocouple. At $40~\mathrm{L\,min^{-1}}$, the bulk velocity at the throat is $U_b = U_t = 13.33~\mathrm{m\,s^{-1}}$, twice the mean inlet velocity $U_{\mathrm{in}} = 6.67~\mathrm{m\,s^{-1}}$, and the Reynolds number based on the equivalent diameter of the throat, $D_h = 6.67$~mm (four times the throat area divided by its perimeter), is $\Rey = U_b D_h/\nu = 95\,160$. The static pressures were measured 54.8~mm upstream of the throat ($P_1$), where the channel height is 20~mm, and 90.5~mm downstream of it ($P_2$), with absolute pressure transmitters (Keller PAA-20SX, ranges 0--5 and 0--3~bar, accuracy $\pm 0.05\,\%$ of full scale) with a response rate of 1~kHz. Their signals were acquired at 100~kHz with a National Instruments data-acquisition system triggered by the high-speed camera, and only time averages are used.

The inception cavitation number is defined as $\sigma_i = (P_1 - p_v)/(\frac{1}{2}\rho U_{\mathrm{in}}^2)$, where $P_1$ is the time-averaged inlet pressure of a recording at inception, $p_v = 2.81$~kPa is the vapor pressure at $23\,^{\circ}\mathrm{C}$ and $\rho = 997.5~\mathrm{kg\,m^{-3}}$. The values reported are means over 2--6 recordings per surface. The overall pressure loss is characterized by $K_{\mathrm{loss}} = (P_1 - P_2)/(\frac{1}{2}\rho U_t^2)$. The throat has no pressure tap, so the throat pressure is estimated from the one-dimensional inviscid relation $P_t = P_1 + \frac{1}{2}\rho\,(U_{\mathrm{in}}^2 - U_t^2)$. This estimate neglects losses and boundary-layer blockage in the converging section, but that section is identical and unetched on all inserts, so its error is common to all surfaces and the differences between them are preserved.

\subsection{High-speed imaging}\label{sec:imaging}

The test section was imaged from the side in back-illumination (figure~\ref{fig:expsetup}\textit{b}), giving shadowgraphs of the streamwise--wall-normal plane at mid-span in which vapor and nuclei appear dark on a bright background. A Phantom TMX~6410 high-speed camera was used for recording at 102\,247~frames~s$^{-1}$ with an exposure time of 1.3~$\mu$s. Illumination was provided by a continuous LED source (GSVITEC LT~Max), collimated and focused onto the test section with a condenser lens and an achromat ($L_1$ and $L_2$ in figure~\ref{fig:expsetup}\textit{b}). Composite images (figures~\ref{fig:traveling_bubble} and \ref{fig:inception_cycle}) are minimum-intensity projections of selected frames, computed in ImageJ \citep{Schneider2012}, so that bubbles and vapor from different instants appear in one image (appendix~\ref{appA}).

\begin{figure}
  \centering
  \includegraphics[width=0.85\textwidth]{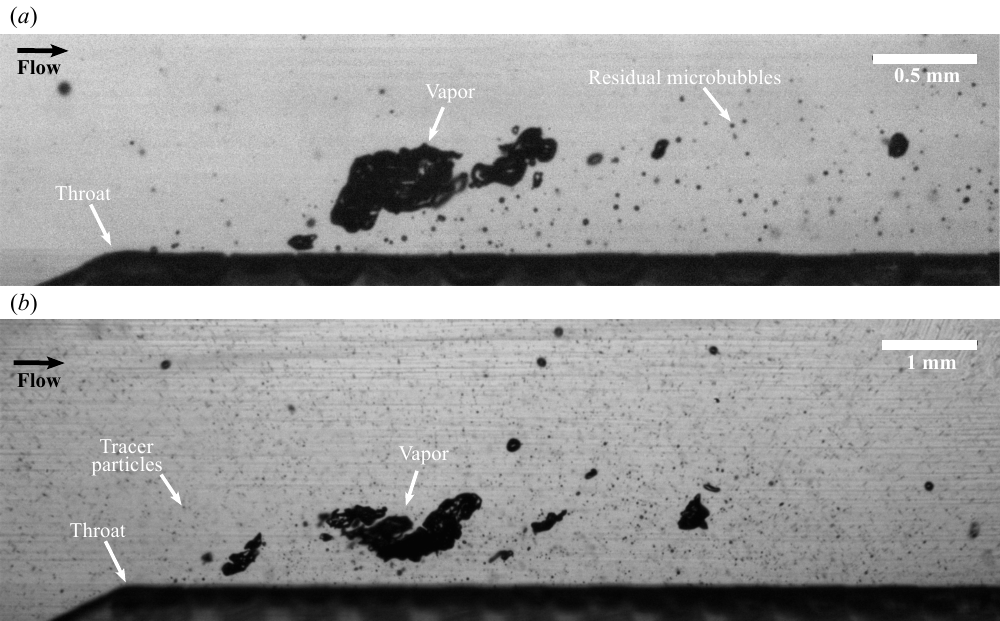}
  \caption{Raw high-speed frames on Rough~4 at inception, shown as recorded with a linear gray scale. (\textit{a}) Nuclei-tracking recording at $3.75~\mu$m~px$^{-1}$, without tracer particles, showing a vapor cloud near the throat and the residual microbubbles downstream of it. (\textit{b}) Velocity-field recording at $8.28~\mu$m~px$^{-1}$, seeded with tracer particles, in which the vapor lies above the wall downstream of the throat. The dark band at the bottom of each frame is the wall.}
  \label{fig:rawframes}
\end{figure}

Two magnifications were used, in separate recordings at the same operating condition. For the tracking of nuclei, an 80/400~mm achromat pair gave a nominal fivefold magnification and a calibrated resolution of $3.75~\mu$m~px$^{-1}$, over a $4.8 \times 1.2$~mm field ($1280 \times 320$~px) that includes the throat corner, and no tracer particles were added. For the velocity fields, a 75/150~mm achromat pair gave a nominal twofold magnification and a calibrated resolution of $8.28~\mu$m~px$^{-1}$, over a $10.6 \times 3.2$~mm field ($1280 \times 384$~px), and the flow was seeded with silver-coated hollow glass spheres (Dantec Dynamics S-HGS) of 10~$\mu$m mean diameter. Their response time, $\tau_p = \rho_p d_p^2/(18\mu) \approx 8~\mu$s for a density of about $1.4~\mathrm{g\,cm^{-3}}$, is short compared with the time scale of the near-wall reverse flow, about 0.13~ms (the reverse-flow layer thickness divided by its peak reverse velocity). It is, however, comparable to the time scale of the steepest mean shear, about 20--40~$\mu$s, so the smallest-scale fluctuations are attenuated on all surfaces. Each recording was registered to its own throat corner and wall. The alignment uncertainty between the two datasets is about $0.003H$, so they are compared statistically rather than event by event. Typical raw frames at the two magnifications are shown in figure~\ref{fig:rawframes}.

For the qualitative observation of figure~\ref{fig:surface_cavity}(\textit{a}), the same Venturi was also imaged in side view at inception with high-speed synchrotron X-ray phase-contrast imaging at beamline 32-ID-B of the Advanced Photon Source (Argonne National Laboratory). The arrangement follows \citet{Zhang2020}, but without the fast and slow shutters used there, so that the images form a continuous time-resolved sequence with a frame interval of 3.62~$\mu$s, at $1.56~\mu$m~px$^{-1}$ (a 2~mm field across 1280~px). X-rays traverse the full span, so they show bubbles inside the surface cavities, which are hidden from the optical view. As for the velocity fields, the flow was seeded with silver-coated hollow glass spheres.

\subsection{Nuclei tracking}\label{sec:tracking}

Nuclei were detected and tracked in the side-view images with TrackMate \citep{Tinevez2017} (appendix~\ref{appA}), and each detection was characterized by its projected area $a$ and equivalent diameter $d_{eq} = (4a/\pi)^{1/2}$. Positions are measured from the throat corner ($x$, positive downstream) and from the wall ($y$) of each recording. A track is classified as upstream-migrating when its net streamwise displacement is at least 0.08~mm against the flow and at least 65\,\% of its steps are directed upstream. The full criteria are given in appendix~\ref{app:tracking}.

\subsection{Velocity fields and vapor masking}\label{sec:piv}

Velocity fields were computed with the optical-flow method RAFT \citep[recurrent all-pairs field transforms,][]{Teed2020}, using its implementation in the Computer Vision Toolbox of MATLAB R2025a, pretrained on the synthetic Kubric dataset \citep{Greff2022} and not retrained for the present images. MATLAB R2025a was also used for the vapor detection and the analysis of the velocity fields. RAFT was applied directly to the raw grayscale images and returns a dense displacement field, one vector per pixel, between two consecutive images. Displacements were converted to velocity with the calibrated resolution of 8.28~$\mu$m~px$^{-1}$ and the frame interval of 9.78~$\mu$s. Each surface was recorded for 0.58~s (59\,095 images at 102\,247~frames~s$^{-1}$), which spans about 770 convective times $H/U_b$, so that each record contains many passages of the largest eddies.

Within vapor, the optical-flow velocity follows the moving vapor image rather than the liquid, so vapor was detected and removed in every image before computing statistics. For each surface, the background image was taken as the median of 50 images spanning the recording and subtracted from each image, and pixels more than 20 gray levels (of 255) darker than the background were classified as vapor. Regions smaller than 20~px$^2$ were discarded, and holes in the remaining regions were filled. A velocity field was masked wherever either of its two images contained vapor, and regions larger than 50~px$^2$ were dilated by 3~px to exclude the biased velocities at their edges. The profiles are insensitive to this dilation between 0 and 8~px. Statistics were then computed from the liquid samples only, without interpolating across vapor, and heights at which fewer than 30\,\% of the frames contain liquid were excluded.

The vapor occupancy $\alpha$, the fraction of images in which a point is vapor, was computed from the masks of the individual images, without combining image pairs or dilating. Uncertainties of the percentiles of its streamwise distribution (\S\ref{sec:location}) are 95\,\% block-bootstrap confidence intervals (5000 resamples).

Profiles were extracted at eleven streamwise stations between $x = 0.5$ and 4.5~mm. Each station is a single column of the velocity field, 8.3~$\mu$m wide, and its statistics are time averages over the full record. At each station, the wall was located as the first row at which liquid motion begins, identified from the noise floor of the velocity fluctuations. This step was needed because, in some recordings, part of the wall face was visible, the wall not being exactly parallel to the line of sight. The wall position is uncertain to $\pm 0.002H$. Statistical convergence was checked with a split-half test, in which statistics computed separately from the first and second halves of each record are compared, and their difference is shown as the error bar. The accuracy of the velocity fields was estimated from the residual displacement between images matched with the measured field \citep{Sciacchitano2013,Wieneke2015}, as described in appendix~\ref{appB}. For $x/H \geq 0.15$, the bias of the near-wall mean velocity estimated in this way is at most $0.04\,U_b$ on the smooth surface, Rough~1, Rough~2 and Rough~4, and at most $0.07\,U_b$ on Rough~3 and $0.08\,U_b$ on Rough~5. Frames free of vapor near the station are used in \S\ref{sec:flow} to test whether vapor alters the near-wall flow. For each station, the region considered extends from the throat to 0.5~mm downstream of the station, below $y/H = 0.1$. A frame is taken as vapor-free if this region contains no vapor patch larger than 50~px$^2$ and vapor covers less than 0.2, 1 or 3\,\% of its area. In these frames, no vapor upstream of the station can drive the flow measured there. All three vapor fraction thresholds give the same result.

\FloatBarrier

\section{Results}\label{sec:results}
\colorlet{orange}{black}

\subsection{Inception delay}\label{sec:delay}

In the initial runs, the rough surfaces incepted at higher flow rates than the smooth surface. This indicated a delay in cavitation inception, but the surfaces were then compared at different Reynolds numbers. With the flow rate held fixed in the main experiments, the roughness can affect inception only through the system pressure at which it occurs. Table~\ref{tab:inception} expresses this pressure in two equivalent ways: as the inception cavitation number $\sigma_i$, which compares the inlet pressure at inception with the vapor pressure, and as the one-dimensional estimate of the throat pressure $P_t$, the absolute pressure near which cavitation first appears. A lower value of either indicates a delayed inception. The table also lists the measured inlet and outlet pressures and the loss coefficient $K_{\mathrm{loss}}$, which is used below to test whether the roughness changes the overall pressure drop.

On the smooth surface, inception occurs at $\sigma_i = 4.07$. On the structured rough surfaces the pressure must be lowered further before sustained cavitation appears: $\sigma_i$ falls to 3.86 on Rough~1 ($k/d = 0.08$) and to 3.33--3.46 on Rough~2--4 ($0.141 \leq k/d \leq 0.444$), a reduction of up to 18\,\% relative to the smooth surface. Expressed as the one-dimensional throat pressure, inception requires $P_t \approx 10$--$13~\mathrm{kPa}$ on Rough~2--4, compared with $26.5~\mathrm{kPa}$ on the smooth surface. At the largest roughness, Rough~5 ($k/d = 0.72$), $\sigma_i$ rises again to 3.79, so the delay is smaller. The six surfaces therefore fall into two groups: the smooth surface with the weakly delayed Rough~1 and Rough~5 ($\sigma_i = 3.79$--$4.07$), and delayed inception on Rough~2--4 ($\sigma_i = 3.33$--$3.46$), referred to below as the delayed surfaces. To check that this grouping is not an artifact of measurement scatter, we compare the gaps with the uncertainties. The expanded uncertainty of $\sigma_i$ is $\pm 0.12$--$0.15$ (table~\ref{tab:inception}). The closest surfaces of the two groups, Rough~3 and Rough~5, differ by 0.33, whereas the uncertainty of this difference is only $\sqrt{0.14^2 + 0.13^2} \approx 0.19$. Rough~1 lies marginally below the smooth surface (by 0.21, close to the uncertainty of 0.20 on that difference) and 0.40 above the delayed surfaces. Rough~1 therefore belongs to the weakly delayed group, but its small delay makes it a borderline case. Rough~5 lies 0.28 below the smooth surface, beyond the uncertainty of 0.19 on that difference, so its delay is smaller than on Rough~2--4 but resolved. The delay is thus non-monotonic in $k/d$: it is largest at intermediate roughness and smaller at the largest, and any explanation must account for both.

A rough wall could affect inception simply by altering the pressure distribution along the Venturi, since roughness can increase wall friction and thicken the boundary layer \citep{Jimenez2004,WuPiomelli2018}. Such a change would appear in the loss coefficient $K_{\mathrm{loss}}$, the total pressure drop across the Venturi scaled by the throat dynamic pressure. The loss coefficient varies by about 6\,\% across all six surfaces ($K_{\mathrm{loss}} = 0.300$--$0.318$), slightly higher on the two roughest surfaces, but shows no relation to $\sigma_i$: Rough~4 and Rough~5 have nearly the same loss coefficient, yet Rough~4 is strongly delayed and Rough~5 only slightly. The delay is thus not explained by a change in the global pressure distribution, which points instead to more local effects near the wall.

A recessed-wall control experiment confirms that the delay is caused by the roughness rather than by the material removed. Etching removes material from the diverging wall and so slightly enlarges the channel, which could by itself alter the local pressure. This control separates that geometric effect from the effect of the cavities. For Rough~2 and Rough~4, the volume removed by etching, measured with the confocal microscope, was converted into a uniform recess of the diverging wall, 2.3 and 11.1~$\mu$m deep respectively, and Venturi inserts with this smooth recessed wall were printed and tested. In this protocol, a delayed inception appears as a higher flow rate at inception. Both incepted at $47.5 \pm 0.5~\mathrm{L\,min^{-1}}$, the same flow rate as the smooth surface, whereas the etched Rough~2 and Rough~4 surfaces incepted at 51 and $54.5~\mathrm{L\,min^{-1}}$ under the same protocol. A uniform recess of this depth changes the channel height by less than 0.2\,\% and does not reproduce the delay, so the delay requires the roughness itself.
 
These measurements do not, however, show how the cavities act. Since inception requires both a nucleus and a sufficiently strong low-pressure event, the cavities can delay it in two ways: by reducing the supply of nuclei, or by changing the near-wall flow that exposes the nuclei to low pressure. To distinguish between them, we first describe, from the high-speed imaging, how cavitation is initiated and sustained on these surfaces and where its nuclei come from. We then test the first possibility by comparing the supply of nuclei across the surfaces, and the second by locating where vapor forms and measuring the near-wall flow.

\begin{table}
    \centering

  \begin{tabular}{lcccccc}
      \hline
      Surface & $k/d$ & $P_1$ (bar) & $P_2$ (bar) & $\sigma_i$ & $P_t$ (kPa) & $K_{\mathrm{loss}}$ \\
      \hline
      Smooth   & $0.030$ & $0.930$ & $0.662$ & $4.07 \pm 0.14$ & $26.5$ & $0.302$ \\
      Rough 1  & $0.080$ & $0.883$ & $0.612$ & $3.86 \pm 0.15$ & $21.8$ & $0.306$ \\
      Rough 2  & $0.141$ & $0.783$ & $0.514$ & $3.41 \pm 0.12$ & $11.8$ & $0.304$ \\
      Rough 3  & $0.267$ & $0.795$ & $0.529$ & $3.46 \pm 0.14$ & $13.0$ & $0.300$ \\
      Rough 4  & $0.444$ & $0.765$ & $0.486$ & $3.33 \pm 0.12$ & $10.0$ & $0.315$ \\
      Rough 5  & $0.720$ & $0.869$ & $0.587$ & $3.79 \pm 0.13$ & $20.4$ & $0.318$ \\
      \hline
  \end{tabular}

  \caption{Pressures at inception for the six surfaces at the fixed flow rate of $40~\mathrm{L\,min^{-1}}$ ($\Rey = 95\,160$): the inlet and outlet static pressures $P_1$ and $P_2$, the inception cavitation number $\sigma_i$, the one-dimensional throat pressure $P_t$ and the loss coefficient $K_{\mathrm{loss}}$, as defined in \S\ref{sec:operating}. A lower $\sigma_i$ or $P_t$ indicates a delayed inception. Values are means over 2--6 recordings per surface. The uncertainties of $\sigma_i$ are expanded uncertainties (95\,\% confidence) combining the scatter between recordings, the transducer accuracy and the fluctuation of the flow rate.}
  \label{tab:inception}
\end{table}


\subsection{Inception cycle}\label{sec:inception_mechanisms}

The high-speed imaging shows the same sequence of events on all six surfaces. Cavitation is triggered by bubbles advected close to the wall, and it is then sustained by the nuclei that these events leave behind. The structured rough surfaces add a further supply of nuclei, grown in the etched cavities. The three stages are described in turn below.

\subsubsection*{Triggering by near-wall traveling bubbles}

\begin{figure}[H]
  \centering
  \includegraphics[width=0.85\textwidth]{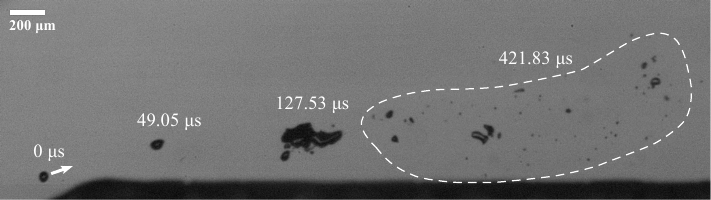}
  \caption{Triggering by a near-wall traveling bubble on Rough~4, shown as a minimum-intensity composite of four frames. The dashed outline encloses the residual microbubbles left after the collapse.}
  \label{fig:traveling_bubble}
\end{figure}

Inception did not start from the small nuclei carried by the bulk flow. In the imaging, most of these nuclei were advected through the throat region without cavitating and were not trapped in the slow near-wall flow. The few that were trapped did not grow enough for their collapse to leave behind several new nuclei. Cavitation was instead initiated by bubbles several times larger than these nuclei, advected close to the wall from upstream of the throat (figure~\ref{fig:traveling_bubble}). As such a bubble passes the region of minimum pressure, it grows rapidly, cavitates and collapses downstream of the throat, leaving behind a cloud of residual microbubbles, often several of them close to the wall. These events are intermittent and do not by themselves sustain cavitation, but they provide the first seed nuclei for the cycle described next. \citet{Ram_Agarwal_Katz_2020} similarly observed traveling-bubble cavitation acting as the initial trigger for patchy attached cavitation, with free-stream nuclei that attached to the surface near the point of minimum pressure seeding the subsequent events.

\subsubsection*{Recycling of nuclei}

Inception becomes sustained once the residual microbubbles start a feedback loop near the wall. A fraction of them, rather than being swept downstream, migrates upstream within the low-momentum near-wall region. The adverse pressure gradient in the diverging section drives them upstream, and near-wall reverse flow, where present (\S\ref{sec:flow}), also carries them. \citet{Ram_Agarwal_Katz_2020} showed that such microbubbles can also grow by diffusion of non-condensable gas during this migration. Here, however, the migration lasts only a few milliseconds, whereas diffusion would take of order $0.1$~s to double their radius, so the recycled nuclei cavitate when they meet a sufficiently strong local pressure drop rather than after growing by diffusion. Each new cavitation event collapses into a fresh set of microbubbles and the cycle repeats, so that inception is sustained largely by recycled nuclei rather than by those carried in by the flow.

\begin{figure}[H]
    \centering
    \includegraphics[width=0.95\textwidth]{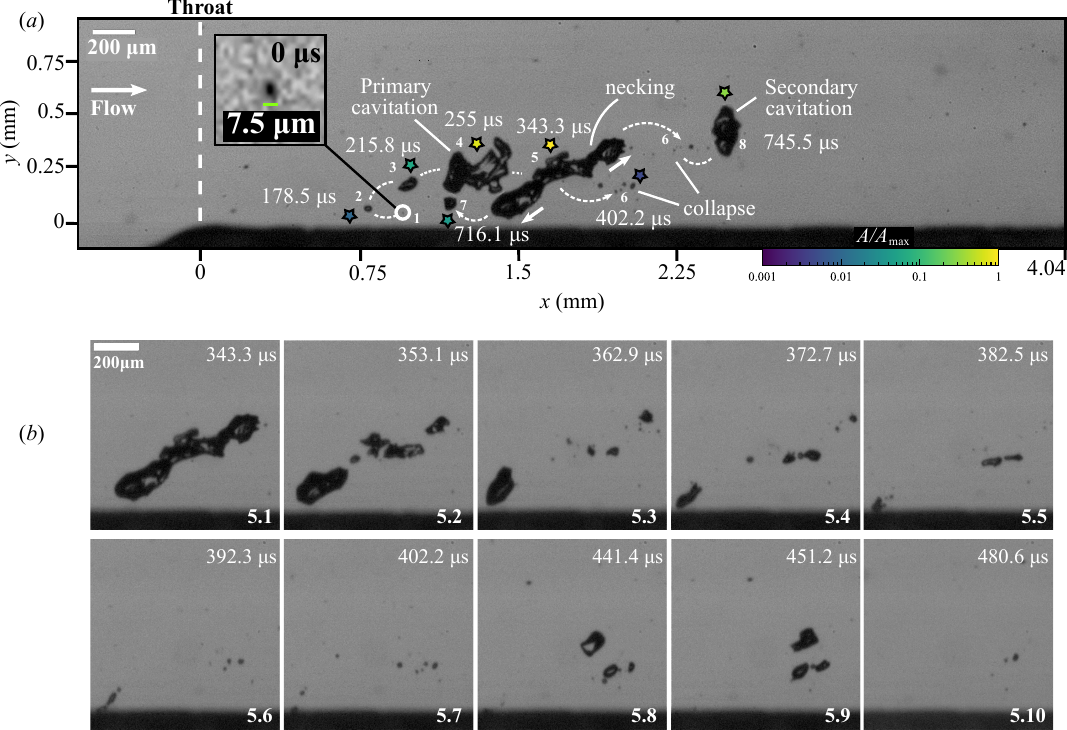}
    \caption{One inception cycle on Rough 4. (\textit{a}) Minimum-intensity composite of the stages of the cycle, numbered 1--8. The color of each marker from stage~2 onward gives the projected area $A$ of the cavity relative to its maximum $A_{\rm max}$, reached at stage~5, on a logarithmic scale. Inset: the initial nucleus (stage~1). (\textit{b}) Frames 5.1--5.10 of stage~5, showing the elongated cavity collapsing, fragmenting and leaving a cluster near the wall.}
    \label{fig:inception_cycle}
\end{figure}

Figure~\ref{fig:inception_cycle} follows one such cycle on Rough 4. A near-wall nucleus of diameter about $7.5~\mu$m (stage~1, $t = 0$) first grows gradually over about $216~\mu$s, while it moves slightly upstream and then lifts about 0.25~mm away from the wall (stages~2 and~3). It then grows explosively, its area increasing by more than an order of magnitude within about $40~\mu$s, and reaches primary cavitation at $t = 255~\mu$s (stage~4). The slow initial growth is still far too fast to be explained by gas diffusion. It is consistent instead with quasi-static expansion as the nucleus enters lower pressure, until the pressure falls below its critical value and the growth becomes explosive \citep{Franc2004}. In this cycle, explosive growth thus begins only after the nucleus has moved away from the wall, as \citet{Russell2026} also observed for nuclei that leave stalled near-wall fluid. The cavity then elongates in two directions ($t = 343.3~\mu$s, stage~5) and develops necks at two locations along its length. On collapse ($t = 402.2~\mu$s, stage~6) it fragments into three microbubble clusters, consistent with the two necks (figure~\ref{fig:inception_cycle}\textit{b}). The cluster closest to the wall migrates back upstream along the wall (stage~7, $t = 716.1~\mu$s) and becomes the nucleus of the next cavitation event, closing the cycle. The cluster further from the wall is either swept downstream or triggers a secondary cavitation event (stage~8, $t = 745.5~\mu$s), which also collapses and is swept away.

\subsubsection*{Surface nuclei on the structured surfaces}

\begin{figure}[H]
  \centering
  \includegraphics[width=0.95\textwidth]{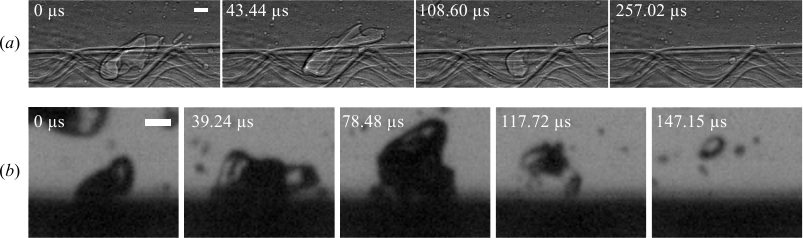}
  \caption{Growth of a bubble from a surface cavity on Rough~4. (\textit{a}) Synchrotron X-ray side view, projected through the span, showing a bubble that grows within a surface cavity and then leaves it. The small features in the flow are seeding particles or nuclei, which cannot be distinguished at this size. (\textit{b}) Optical side view of a bubble growing out of the wall ($t = 0$--$78.48~\mu$s), lifting off and collapsing ($t = 117.72~\mu$s), and leaving residual nuclei ($t = 147.15~\mu$s). Scale bars: $50~\mu$m.}
  \label{fig:surface_cavity}
\end{figure}

The two most deeply etched surfaces, Rough~4 and Rough~5, provide a third source of nuclei, absent on the other surfaces. Bubbles were observed to grow within the etched cavities and then detach into the near-wall flow (figure~\ref{fig:surface_cavity}). The synchrotron X-ray images show a bubble growing inside a cavity and leaving it (figure~\ref{fig:surface_cavity}\textit{a}), and the optical side view shows a bubble growing out of the wall, lifting off and collapsing (figure~\ref{fig:surface_cavity}\textit{b}). Such events are rare. Over 20\,000 frames, about 6 were observed on Rough~4 and 14 on Rough~5, and none on Rough~1--3, each lasting 60--120 frames (0.6--1.2~ms), whereas about 3\,700 and 3\,500 nuclei, respectively, migrated upstream along the wall on these two surfaces over the same number of frames (figure~\ref{fig:supply}\textit{a}). \citet{ArndtIppen1968} similarly observed gaseous cavitation in the roughness grooves of a rough-wall boundary layer, with nuclei only occasionally leaving the grooves and few cavitating nuclei near the wall. The bubbles did not sustain cavitation themselves. Their collapse left behind nuclei that joined the cycle of inception, either starting an event or adding to one already under way.

Rough~4 therefore has, if anything, more nuclei available than the smooth surface, yet its inception is the most delayed. In the next subsection, we test whether the main supply of nuclei, those migrating upstream along the wall, differs across the surfaces.

\subsection{Supply of nuclei}\label{sec:supply}

The inception cycle is sustained by nuclei that migrate upstream along the wall toward the throat. To compare this supply across the surfaces, we count on each surface the tracks classified as upstream-migrating with the criteria of \S\ref{sec:tracking}.

\begin{figure}[H]
  \centering
  \includegraphics[width=\textwidth]{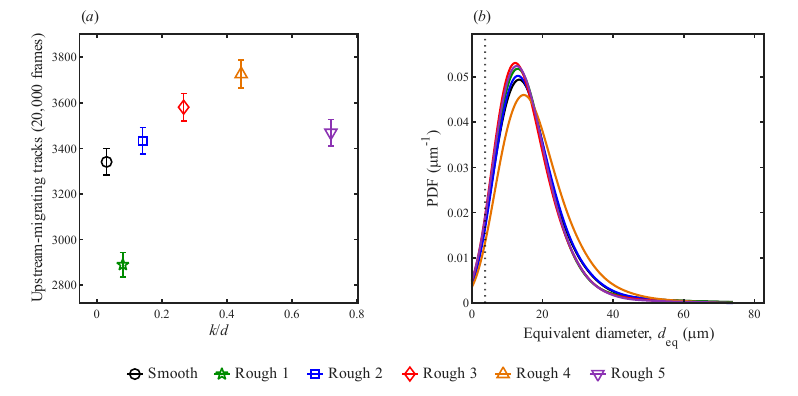}
  \caption{Supply of upstream-migrating nuclei at $\Rey = 95\,160$. (\textit{a}) Number of upstream-migrating tracks over 20\,000 frames versus $k/d$. Error bars show one standard deviation of the count. (\textit{b}) Probability density of the equivalent diameter $d_{eq}$ of the upstream-migrating nuclei (Gaussian kernel, bandwidth 3.75~$\mu$m). The dotted line marks one pixel.}
  \label{fig:supply}
\end{figure}

The supply of nuclei is similar on all surfaces (figure~\ref{fig:supply}\textit{a}). Over 20\,000 frames, the number of upstream-migrating tracks lies between about 2900 and 3700, with no monotonic trend with $k/d$. It is smallest on Rough~1 and largest on Rough~4, so the delayed surfaces (Rough~2--4) receive as many upstream-migrating nuclei as the smooth surface, or more. The size distributions of these nuclei are also nearly identical (figure~\ref{fig:supply}\textit{b}). All peak at $d_{eq} \approx 12$--$15~\mu$m, close to the imaging resolution, and few nuclei exceed $40~\mu$m. Only Rough~4 has a slightly heavier tail, between 20 and $35~\mu$m. Each surface was recorded at its own inception condition, so the delayed surfaces operated at a lower pressure, which would, if anything, favor the production and growth of nuclei. The delay is therefore not caused by a shortage of nuclei near the wall. If the roughness does not change how many nuclei arrive, it must change how they are exposed to low pressure, and we examine this next by locating where vapor forms on each surface.

\subsection{Location of vapor at inception}\label{sec:location}

We obtain the vapor regions at inception from the masks of the individual frames of the velocity recordings (\S\ref{sec:piv}). From these we compute the near-wall vapor occupancy $\alpha(x)$, the time-averaged fraction of the band $0 \leq y/H < 0.05$ occupied by vapor at each streamwise position. Treating $\alpha(x)$ as a distribution along $x$, its 10th, 50th and 90th percentiles mark the upstream edge, the middle and the downstream tail of the region where vapor forms. Since each surface was recorded at its own inception pressure, the amount of vapor differs between surfaces, so we compare where the vapor forms rather than how much of it there is.

\begin{figure}
  \centering
  \includegraphics[width=\textwidth]{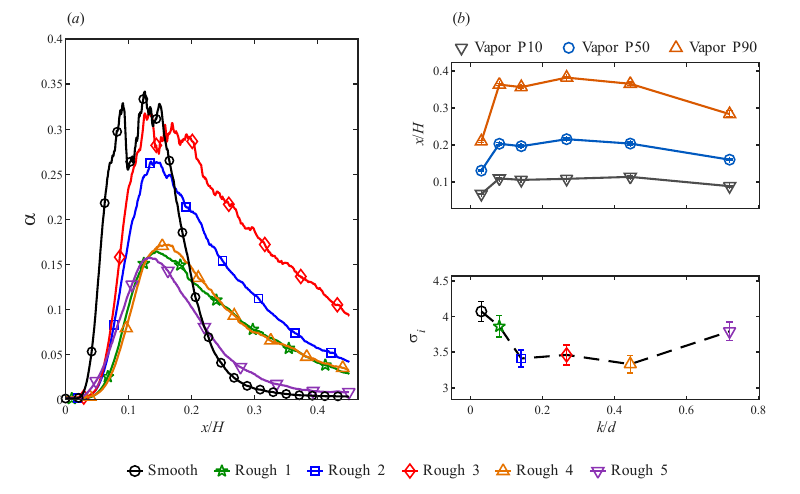}
  \caption{Vapor location at inception. (\textit{a}) Near-wall vapor occupancy $\alpha$, averaged over $0 \leq y/H < 0.05$, versus $x/H$. (\textit{b}) Top: 10th, 50th and 90th percentiles of the streamwise distribution of $\alpha$ versus $k/d$. The 95\,\% block-bootstrap confidence intervals are smaller than the markers. Bottom: $\sigma_i$ with its 95\,\% expanded uncertainty.}
  \label{fig:vaporloc}
\end{figure}

On the smooth surface, vapor appears closest to the throat (figure~\ref{fig:vaporloc}\textit{a}). Its occupancy rises from $x/H \approx 0.03$ and falls almost to zero by $x/H \approx 0.3$. On Rough~1--4 the occupancy rises further downstream and decays slowly, so that vapor is still present at the end of the field ($x/H = 0.45$). The percentiles quantify this displacement (figure~\ref{fig:vaporloc}\textit{b}). On the smooth surface, the upstream edge of the vapor lies at $x/H = 0.068$ and its median at 0.131. On Rough~1--4 the upstream edge lies at $x/H = 0.106$--0.114 and the median at 0.197--0.216, about 0.04 and 0.07--0.09 further from the throat. Rough~5 lies between the two groups (0.089 and 0.161). The 95\,\% confidence intervals are $\pm 0.001$--0.005, so these differences are well resolved. The displacement does not, however, follow $\sigma_i$ exactly. Rough~1 is displaced as far as the delayed surfaces, although its inception cavitation number is only marginally below that of the smooth surface. As shown in \S\ref{sec:flow}, the displacement instead occurs on every surface that has a near-wall reverse-flow layer.

\begin{figure}
  \centering
  \includegraphics[width=\textwidth]{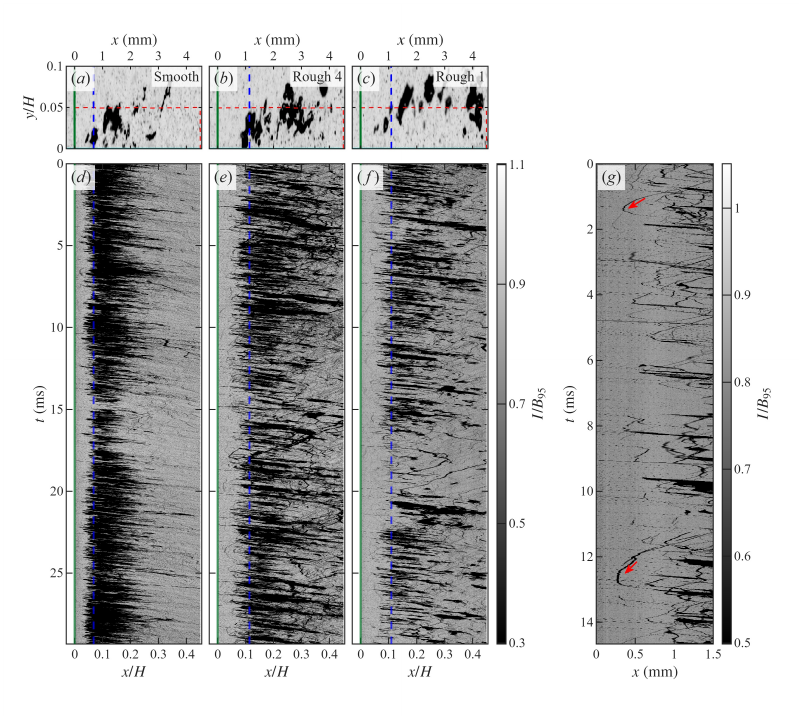}
  \caption{Near-wall vapor and nuclei in space and time. (\textit{a}--\textit{c}) Snapshots of the normalized intensity $I/B_{95}$ on the smooth surface, Rough~4 and Rough~1, rotated so that the wall lies along $y/H = 0$, with the vertical scale stretched about threefold. Red dashed box: analysis band $0 \leq y/H \leq 0.05$. Green line: throat corner. Blue dashed line: upstream edge of the vapor (figure~\ref{fig:vaporloc}\textit{b}). (\textit{d}--\textit{f}) Space--time diagrams of the minimum of $I/B_{95}$ across the band, over 3000 consecutive frames (29.3\,ms). (\textit{g}) Rough~4, from a tracking recording, for $x \leq 1.5$\,mm over 1500 frames, in a window chosen for low vapor near the throat. Arrows mark nuclei moving toward the throat.}
  \label{fig:spacetime}
\end{figure}

The space--time diagrams of figure~\ref{fig:spacetime} show how this displacement develops in time. Each image is normalized by $B_{95}$, the 95th-percentile intensity of each pixel over the recording, which represents the liquid background. A median background is not used because the persistent vapor near the throat on the smooth surface would bias it. In each diagram, a row is one frame, in which the darkest point across the near-wall band is plotted at each $x$, so that vapor appears as dark marks and time runs downward. The windows shown were chosen to contain inception events, and the isolated gray speckle is partly tracer particles. On the smooth surface, vapor occupies the band almost continuously from $x/H \approx 0.03$ and stays within $x/H \lesssim 0.2$. On Rough~4 and Rough~1, the band upstream of $x/H \approx 0.07$--$0.1$ stays free of vapor. Downstream of it, vapor appears in streaks that spread over the rest of the field, more intermittently on Rough~1. A vapor structure crosses the field in about 0.4\,ms, so these diagrams show where vapor resides rather than how it moves.

Figure~\ref{fig:spacetime}(\textit{g}) shows the same band on Rough~4 from a tracking recording, which has no tracer particles, in a window chosen for low vapor near the throat. A nucleus moving through the band traces a slanted line, whose slope gives its speed. The marked nuclei drift toward the throat at about 0.4--0.5\,m\,s$^{-1}$, stop more than about 0.2\,mm ($x/H \approx 0.02$) short of it and are then swept downstream. Such nuclei are rare. Many more traces appear further downstream ($x \approx 0.8$--1.5~mm), where nuclei migrate upstream and turn within the cavitating region. In two tracked 2000-frame windows of the same recording, about a quarter to a third of the upstream-migrating nuclei near the wall reached $x < 1$~mm, but only 1--3\,\% reached $x < 0.5$~mm, and some turned back downstream as the marked nuclei do. On Rough~1--4, vapor thus forms further downstream, and the nuclei migrating along the wall rarely reach the region closest to the throat. The next subsection examines the near-wall flow and where the vapor lies relative to it.

\subsection{Near-wall flow}\label{sec:flow}

To examine the flow in which vapor forms, we extract liquid-phase statistics from the velocity fields at eleven streamwise stations between $x/H = 0.05$ and 0.45, using all frames of each recording and excluding vapor (\S\ref{sec:piv}). We report the mean streamwise velocity $\bar{u}$ and the in-plane fluctuation energy $k_{2D} = \frac{1}{2}(\overline{u'^2} + \overline{v'^2})$, both normalized by the bulk throat velocity $U_b$, together with the vapor occupancy $\alpha$. Two quantities describe the structure of the near-wall flow. The shear-layer height $y_{0.5}$ is the height at which $\bar{u}$ lies halfway between its near-wall minimum $\bar{u}_{\min}$, taken below $y/H = 0.03$, and the maximum of the profile $\bar{u}_{\max}$. This is analogous to the position $y_{0.5}$ in a plane mixing layer, at which the mean velocity lies midway between those of the two streams \citep{Pope2000}, and it remains defined when the near-wall flow is reversed. The thickness of the reverse-flow layer, $\delta_r$, is the highest point at which $\bar{u} < -0.02\,U_b$. This threshold is several times larger than the statistical uncertainty of the near-wall mean velocity, typically $0.003\,U_b$, but comparable to its possible bias (appendix~\ref{appB}), so the layer was also tested with larger thresholds. With thresholds of 0.01, 0.02, 0.03 and $0.05\,U_b$, a reverse-flow layer appears at most stations on Rough~1--4 and at no more than one station on the smooth surface and Rough~5. To detect reverse flow that the mean hides, we also compute the reverse-flow fraction $\chi$, the fraction of the liquid samples in which the instantaneous streamwise velocity is below the same threshold, $u < -0.02\,U_b$, averaged over $0 \leq y/H < 0.005$. This band is the thinnest reverse-flow layer measured where the layer first appears, so it lies within the layer wherever the layer exists, and it is where the nuclei migrate along the wall.

\begin{figure}
  \centering
  \includegraphics[width=\textwidth]{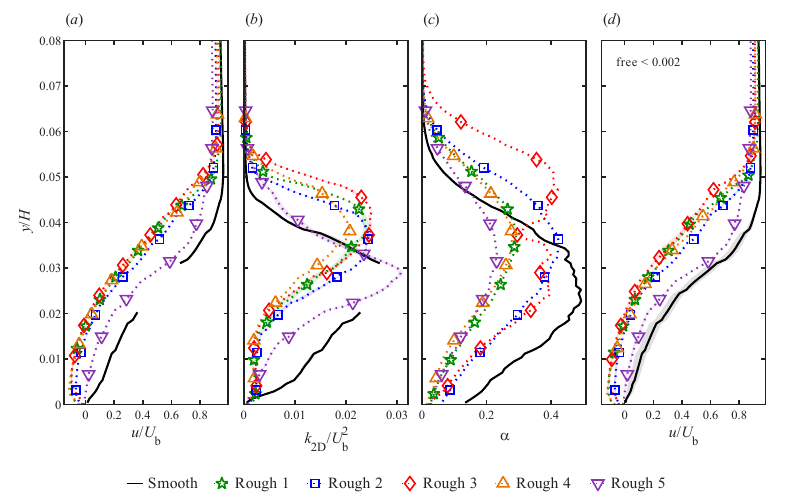}
  \caption{Liquid-phase profiles at $x/H = 0.15$. (\textit{a}) Mean streamwise velocity, (\textit{b}) in-plane fluctuation energy, (\textit{c}) vapor occupancy $\alpha$, (\textit{d}) mean streamwise velocity in vapor-free frames (\S\ref{sec:piv}), using the strictest vapor-fraction threshold of 0.2\,\% (`free $<$ 0.002'), 223--6819 frames per surface. Gaps in (\textit{a}) and (\textit{b}) are heights with valid liquid samples in fewer than 30\,\% of the frames. Panel (\textit{c}) is computed from all images and has no gaps. Shading shows the difference between the two halves of each record.}
  \label{fig:profiles}
\end{figure}

Figure~\ref{fig:profiles} compares the profiles at $x/H = 0.15$, within the region where vapor forms on all surfaces. The surfaces separate into two groups. On Rough~1--4 the mean flow near the wall is reversed (figure~\ref{fig:profiles}\textit{a}) and the shear layer is lifted away from the wall. Close to the wall, where the nuclei travel, the fluctuations are much weaker on Rough~1--4 (figure~\ref{fig:profiles}\textit{b}). At $y/H = 0.02$, the fluctuation energy is $k_{2D}/U_b^2 = 0.004$--0.007 on Rough~1--4, a sixth to a third of its value on the smooth surface (0.0225), with Rough~5 in between (0.015). On the smooth surface and Rough~5 there is no mean reverse flow, the shear layer lies closer to the wall, and strong fluctuations reach the wall. On Rough~5, which has the deepest cavities, the near-wall mean velocity stays above the reverse-flow threshold ($\min \bar{u} \approx -0.01\,U_b$). The reverse-flow fraction varies gradually across the surfaces rather than dividing them into two groups. On the smooth surface the flow is reversed in at most 4\,\% of the liquid samples, near $x/H = 0.1$--0.15. On Rough~5 it is reversed in up to 32\,\% near $x/H = 0.15$--0.2, where its mean near-wall velocity is close to zero, so the flow there is close to separation and reverses intermittently. On Rough~1--4 it is reversed in more than half of the samples from $x/H = 0.15$ downstream. Splitting each record into two halves changes the fractions by at most 4 percentage points, much less than the gaps of 20 points or more between the smooth surface, Rough~5 and Rough~1--4, so the fractions are statistically stable. The ranking of the surfaces, from the smooth surface through Rough~5 to Rough~1--4, is also unchanged when the band is doubled to $0 \leq y/H < 0.01$ or when the wall position is moved by its uncertainty of $0.002H$. Counting every negative velocity, however small, gives larger fractions, up to 28\,\% on the smooth surface, but there the mean near-wall velocity is only slightly positive, so small measurement errors alone push many samples below zero. Upstream of $x/H = 0.15$, Rough~1 differs from Rough~2--4: its flow begins to reverse closer to the throat, in 62\,\% of the samples at $x/H = 0.1$ compared with 28--45\,\% on Rough~2--4.

\begin{figure}
  \centering
  \includegraphics[width=\textwidth]{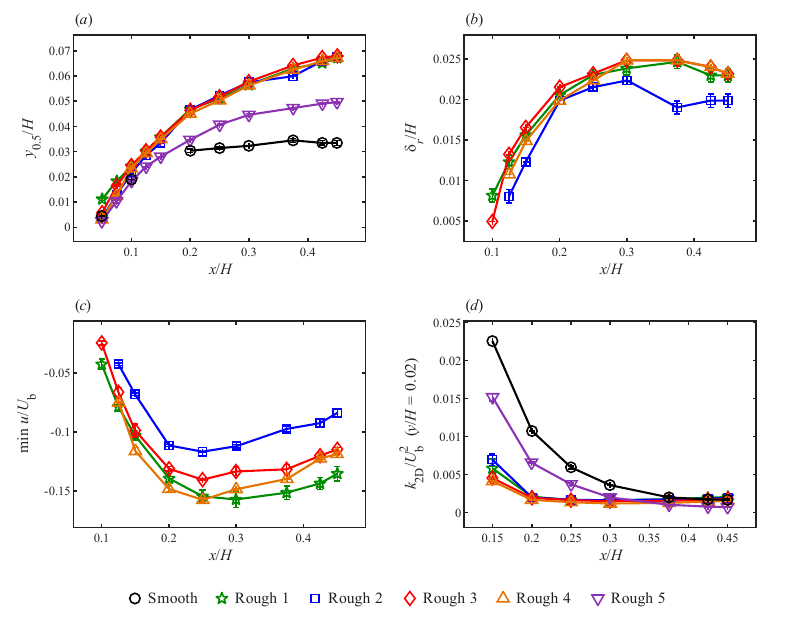}
  \caption{Streamwise development of the near-wall flow. (\textit{a}) Shear-layer height $y_{0.5}/H$, (\textit{b}) reverse-flow-layer thickness $\delta_r/H$, (\textit{c}) minimum mean velocity below $y/H = 0.03$, shown only where a reverse-flow layer exists, and (\textit{d}) fluctuation energy $k_{2D}/U_b^2$ at $y/H = 0.02$ for $x/H \geq 0.15$. Upstream of this station, $y/H = 0.02$ lies within the shear layer. Missing smooth-surface points in (\textit{a}) are stations excluded by vapor. Error bars show the difference between the two halves of each record.}
  \label{fig:streamwise}
\end{figure}

The streamwise development of the near-wall flow also differs between the two groups (figure~\ref{fig:streamwise}). On Rough~1--4 the reverse-flow layer thickens from $\delta_r/H \approx 0.005$--0.011 where it first appears ($x/H = 0.1$--0.125) to 0.022--0.025 by $x/H = 0.3$, and keeps about this thickness further downstream (figure~\ref{fig:streamwise}\textit{b}). The reverse flow strengthens between $x/H = 0.1$ and 0.3, reaching $\min \bar{u} = -0.12$ to $-0.16\,U_b$ at $x/H = 0.2$--0.3, and weakens slightly beyond $x/H = 0.3$ (figure~\ref{fig:streamwise}\textit{c}). The shear layer rises continuously, with $y_{0.5}/H$ increasing from 0.024 at $x/H = 0.1$ to 0.068 at $x/H = 0.45$ (figure~\ref{fig:streamwise}\textit{a}). Near-wall vapor occupancy on Rough~1--4 persists downstream (figure~\ref{fig:vaporloc}\textit{a}). On the smooth surface, the shear layer stops rising beyond $x/H = 0.2$, where $y_{0.5}/H$ remains at 0.030--0.034, and its near-wall fluctuations decay until they match those of Rough~1--4 at $x/H \approx 0.375$ (figure~\ref{fig:streamwise}\textit{d}). Its near-wall vapor falls almost to zero by $x/H = 0.3$ (figure~\ref{fig:vaporloc}\textit{a}). Rough~5 lies between the two groups. Its shear layer rises more slowly than on Rough~1--4, with $y_{0.5}/H$ increasing from 0.018 at $x/H = 0.1$ to 0.050 at $x/H = 0.45$, and, as on the smooth surface, its near-wall vapor dies out by $x/H \approx 0.35$.

Vapor can itself alter the near-wall flow, since a cavity attached near the wall can drive reverse flow beneath it and displace the shear layer \citep{Franc2004}. To test whether the reverse flow is caused by the roughness or by the vapor, we recomputed the mean velocity from frames that contain no vapor between the throat and the measurement station (figure~\ref{fig:profiles}\textit{d}). Between 223 and 11\,775 such frames are available per surface and station, except on Rough~3 at $x/H = 0.3$, where there are only 3--16. Near the wall, the vapor-free profiles nearly coincide with those over all frames (figure~\ref{fig:profiles}\textit{a}). The reverse flow persists in them, with $\bar{u} \approx -0.12$ and $-0.13\,U_b$ on Rough~1 and Rough~4 at $x/H = 0.15$, and the shear layer remains lifted. The smooth surface shows no reverse flow with or without vapor. The reverse-flow layer and the lifted shear layer are therefore properties of the flow over the roughened wall, not a consequence of cavitation.

On Rough~1--4, the reverse-flow layer lies between the wall, along which the nuclei migrate upstream, and the lifted shear layer, where the fluctuations are strongest. The near-wall fluctuations there remain a sixth to a third of the smooth-surface value over $0.15 \leq x/H \leq 0.25$ (figure~\ref{fig:streamwise}\textit{d}), and the vapor forms above the reverse-flow layer rather than on the wall (figure~\ref{fig:profiles}\textit{c}). On the smooth surface the vapor occupancy peaks at $y/H \approx 0.023$ and extends down to the wall, whereas on Rough~1--4 it peaks higher, at $y/H \approx 0.035$--0.047, and falls toward the wall. Within 0.1~mm of the wall ($y/H < 0.01$), vapor is present in 21\,\% of the images on the smooth surface, compared with 4--11\,\% on Rough~1--4. To check that this is not simply because the smooth surface carries more vapor, we divide the near-wall occupancy of each surface by its own average over the near-wall band of \S\ref{sec:location}, the 0.5~mm above the wall ($0 \leq y/H < 0.05$). This ratio is about two thirds on the smooth surface (0.63) but only a quarter to a third on Rough~1--4 (0.23--0.37), so the vapor on Rough~1--4 sits further from the wall regardless of its amount. The contrast holds for $0.125 \leq x/H \leq 0.2$ and for wall shifts of up to $0.002H$. On Rough~5 the vapor also stays away from the wall at $x/H = 0.15$, where it is present within 0.1~mm of the wall in 5\,\% of the images and the ratio is about a third. Nearer the throat ($x/H \leq 0.125$), however, its ratio is larger than on Rough~1--4.

In the mean, the flow therefore divides the surfaces into Rough~1--4, which have a reverse-flow layer, and the smooth surface and Rough~5, which do not, although the flow over Rough~5 reverses intermittently. The vapor is displaced downstream on all four surfaces with the layer, and only slightly on Rough~5. The grouping also matches that of $\sigma_i$ on five of the six surfaces. The exception is Rough~1, which has both the reverse-flow layer and the displaced vapor but only a marginal delay (\S\ref{sec:rough1}). The next subsection asks how the reverse-flow layer translates into the lower cavitation number needed for inception.

\subsection{Link between the near-wall flow and inception}\label{sec:link}

Two contributions could make inception on Rough~2--4 require a lower cavitation number. The mean pressure recovers along the diverging section, so vapor that forms further downstream sits where the mean pressure is higher. In addition, the intermittent low-pressure events that the nuclei experience may be weaker.

The first contribution can be estimated from one-dimensional continuity. Along the diverging section the flow area grows and the mean velocity falls, so the mean pressure rises. With one wall diverging at $8^{\circ}$ from the throat, this rise above the throat pressure, scaled by $\frac{1}{2}\rho U_t^2$, is the ideal recovery coefficient $C_p(x) = 1 - [H/(H + x\tan 8^{\circ})]^2$. Since lowering the system pressure at a fixed flow rate shifts all pressures by the same amount, and $\sigma_i$ is scaled by the inlet dynamic pressure with $U_t = 2U_{\mathrm{in}}$, a change $\Delta C_p$ in the local mean pressure at inception corresponds to $\Delta\sigma_i = 4\,\Delta C_p$. For Rough~2--4, the additional recovery corresponds to $\Delta\sigma_i \approx 0.04$--0.05 at the displaced edge of the vapor, 6--7\,\% of the observed reduction of 0.61--0.74 (table~\ref{tab:inception}), and to 0.07--0.09 at the median, at most about 15\,\%. The estimate uses the one-dimensional mean pressure and neglects losses and the low-momentum near-wall flow, both of which reduce the recovery, so it is likely an upper bound. The near-wall pressure beneath the shear layer was not measured.

Most of the delay must therefore come from the second contribution. On the smooth surface, inception occurs while the one-dimensional throat pressure is 26.5~kPa, about 24~kPa above the vapor pressure. The mean flow alone cannot cavitate the nuclei, and inception relies on intermittent low-pressure events, as \citet{Ram_Agarwal_Katz_2020} also concluded for attached cavitation. On Rough~2--4 the throat pressure must be lowered to 10--13~kPa, a further 13.5--16.5~kPa ($0.15$--$0.19$ of $\frac{1}{2}\rho U_t^2$), before inception occurs, and pressure recovery accounts for at most about 2~kPa of this. The low-pressure events that reach the nuclei on these surfaces are thus substantially weaker. This agrees with the near-wall flow. The nuclei migrate along the wall beneath the lifted shear layer, where the fluctuation energy at $y/H = 0.02$ is a sixth to a third of its smooth-surface value for $0.15 \leq x/H \leq 0.25$ (figure~\ref{fig:streamwise}\textit{d}), and the vapor forms above the reverse-flow layer rather than on the wall (figure~\ref{fig:profiles}\textit{c}).

The depth of the low-pressure events needed for inception can be estimated from the mean pressures alone. In the classical inception criterion, cavitation appears when the minimum pressure reaches the vapor pressure, so that $\sigma_i \approx -C_{p,\mathrm{min}}$ \citep{Franc2004}. Following this idea, we define a local cavitation number at the upstream edge of the vapor $x_v$, the 10th percentile of its streamwise distribution (figure~\ref{fig:vaporloc}\textit{b}), as $\sigma_a = [P_t + C_p(x_v)\,\frac{1}{2}\rho U_t^2 - p_v]/(\frac{1}{2}\rho U_t^2)$. Its numerator is the local mean pressure above the vapor pressure given by the one-dimensional recovery, and the subscript $a$ stands for activation. At inception, the deepest low-pressure events must lower the pressure from this local mean to about the vapor pressure, so $\sigma_a$ measures their depth in units of the throat dynamic pressure. Since $P_t = P_1 + \frac{1}{2}\rho(U_{\mathrm{in}}^2 - U_t^2)$, it can also be written as $\sigma_a = \sigma_i/4 - 3/4 + C_p(x_v)$, so its uncertainty is that of $\sigma_i$ divided by four, about $\pm 0.04$. Table~\ref{tab:sigmaa} shows that $\sigma_a \approx 0.29$ on the smooth surface, 0.22--0.24 on Rough~1 and Rough~5, and 0.11--0.14 on Rough~2--4. The low-pressure events that produce vapor on Rough~2--4 are therefore roughly half as deep as on the smooth surface. The nuclei cavitate at a critical pressure a few kilopascals below the vapor pressure rather than exactly at it. This critical pressure is set mainly by their size, which has the same distribution on all surfaces (figure~\ref{fig:supply}\textit{b}), so it shifts $\sigma_a$ by the same small amount on every surface and leaves the difference between them unchanged. Like the recovery estimate, $\sigma_a$ is one-dimensional and does not resolve the near-wall pressure beneath the shear layer.

\subsection{Fluctuation-based inception criterion}\label{sec:criterion}
\begin{figure}
  \centering
  \includegraphics[width=0.5\textwidth]{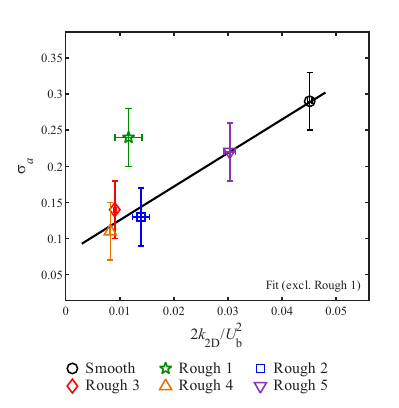}
  \caption{Local cavitation number $\sigma_a$ at the upstream edge of the vapor versus the near-wall fluctuation energy $2k_{2D}/U_b^2$ at $y/H = 0.02$, $x/H = 0.15$. The criterion line is the least-squares fit of (\ref{eq:criterion2}) to all surfaces except Rough~1. Error bars show $\pm 0.04$ in $\sigma_a$ and the difference between the two halves of each record in $k_{2D}$.}
  \label{fig:criterion}
\end{figure}

The link between the depth of these events and the near-wall fluctuations can be made quantitative with the argument used by \citet{ArndtIppen1968} for rough-wall boundary layers. Inception requires the instantaneous pressure at a nucleus to fall to its critical pressure $p_c$. If the excursions that activate the nuclei reach $n$ times the r.m.s.\ pressure fluctuation below the local mean, nuclei are activated, and inception occurs, when,
\begin{equation}
  \bar{p}(x_v) - n\,p'_{\mathrm{rms}} = p_c
  \label{eq:criterion}
\end{equation}
The pressure fluctuations scale with the energy of the velocity fluctuations, $p'_{\mathrm{rms}} = c\,\rho\,k_{2D}$, where $k_{2D}$ equals the mean square of a single velocity component in isotropic turbulence, with $c \approx 0.6$--0.8 at high Reynolds number \citep{Batchelor1951,SchumannPatterson1978}. Subtracting $p_v$ from (\ref{eq:criterion}) and dividing by $\frac{1}{2}\rho U_t^2$, with $U_t = U_b$, gives
\begin{equation}
  \sigma_a = 2N\,\frac{k_{2D}}{U_b^2} + \frac{p_c - p_v}{\frac{1}{2}\rho U_t^2}, \qquad N = nc .
  \label{eq:criterion2}
\end{equation}
Equation~(\ref{eq:criterion2}) differs in two respects from the criterion of \citet{ArndtIppen1968}, whose local inception index scales with the skin-friction coefficient, $\sigma_{il} \approx 16\,C_f$. They inferred the pressure fluctuations from the wall shear stress, whereas here the fluctuation energy is taken at the height of the nuclei. The wall shear is not a suitable measure here, since it vanishes or reverses beneath a reverse-flow layer. \citet{ArndtIppen1968} themselves noted that the activating pressure peaks are probably larger inside the boundary layer than at the wall, and in separated flows even the pressure fluctuations measured at the wall do not account for inception \citep{Katz1984}. They also assumed that inception occurs when the pressure reaches the vapor pressure, whereas here the unknown difference $p_c - p_v$ is retained and appears as the intercept of the relation.
The last term is common to all surfaces. The throat velocity and the vapor pressure are the same because the flow rate and the water temperature are the same. The critical pressure $p_c$ is set mainly by the size and gas content of the nuclei, which are also the same, since all surfaces were tested in the same water and the nuclei have the same size distribution on all of them (figure~\ref{fig:supply}\textit{b}). Nuclei grown in the surface cavities could differ, but they were observed only on Rough~4, the most delayed surface, and on Rough~5, one of the least delayed, and were rare on both (\S\ref{sec:inception_mechanisms}). Equation~(\ref{eq:criterion2}) therefore predicts that $\sigma_a$ varies linearly with the near-wall fluctuation energy, with a single slope for all surfaces. Figure~\ref{fig:criterion} tests this with $k_{2D}$ at $y/H = 0.02$ and $x/H = 0.15$, the station closest to the upstream edge of the vapor at which this height lies below the shear layer. The smooth surface and Rough~2--5 fall on one line (the criterion line) ($N = 4.6$, $R^2 = 0.97$). Rough~1 lies above it and is excluded from the fit, for reasons discussed in \S\ref{sec:rough1}. Including it lowers $R^2$ to 0.63. The fit is controlled mainly by three groups, the smooth surface, Rough~5 and Rough~2--4, since the differences in $\sigma_a$ and $k_{2D}$ within Rough~2--4 are within their uncertainties. Repeating the test at $x/H = 0.20$ and 0.25 gives the same linearity ($R^2 \geq 0.98$), although $N$ itself increases as the fluctuations decay downstream. The linearity is also insensitive to the reference height. At $y/H = 0.015$ the fit gives $R^2 = 0.96$, and at $y/H = 0.025$, where the smooth surface is too often vapor-laden for its liquid statistics to be retained, the four surfaces Rough~2--5 alone give $R^2 = 0.93$.

$N$ is therefore an effective constant for a given reference station rather than a universal one, as expected since inception depends on the tail of the pressure distribution and not only on its r.m.s.\ value \citep{RanKatz1994}. With $c \approx 0.6$--0.8, $n \approx 6$--8. The pressure excursions that trigger inception are thus several times deeper than the r.m.s.\ fluctuation, comparable to the values of about 7 and 10 inferred by \citet{ArndtIppen1968} for a rough-wall boundary layer and by \citet{Rouse1953} for a submerged jet. Rough~1 lies 0.11 above the criterion line at every station and height tested, nearly three times the uncertainty in $\sigma_a$. It incepts at a higher cavitation number than the criterion line gives for its near-wall fluctuation energy. The intercept absorbs the unknown $p_c - p_v$ together with the errors of the one-dimensional mean pressure and of using $k_{2D}$ in place of the full fluctuation energy, and is not interpreted. With five surfaces at one Reynolds number, equation~(\ref{eq:criterion2}) is a consistency test rather than a general law.

\begin{table}
  \begin{center}
  \def~{\hphantom{0}}
  \begin{tabular}{lcccc}
      \hline
      Surface  & $x_v/H$ & $C_p(x_v)$ & $\bar{p}(x_v)$ (kPa) & $\sigma_a$ \\[3pt]
      \hline
      Smooth   & 0.068 & 0.019 & 28.2 & 0.29 \\
      Rough 1  & 0.110 & 0.030 & 24.5 & 0.24 \\
      Rough 2  & 0.106 & 0.029 & 14.4 & 0.13 \\
      Rough 3  & 0.109 & 0.030 & 15.7 & 0.14 \\
      Rough 4  & 0.114 & 0.031 & 12.8 & 0.11 \\
      Rough 5  & 0.089 & 0.025 & 22.6 & 0.22 \\
      \hline
  \end{tabular}
  \caption{Local mean pressure $\bar{p}(x_v)$ and local cavitation number $\sigma_a$ at the upstream edge of the vapor, $x_v$, taken as the 10th percentile of the streamwise vapor distribution (figure~\ref{fig:vaporloc}\textit{b}). The local mean pressure is $\bar{p}(x_v) = P_t + C_p(x_v)\,\frac{1}{2}\rho U_t^2$, where $P_t$ is the one-dimensional throat pressure (table~\ref{tab:inception}) and $C_p$ is the ideal recovery coefficient, and $\sigma_a = [\bar{p}(x_v) - p_v]/(\frac{1}{2}\rho U_t^2)$, with $\frac{1}{2}\rho U_t^2 = 88.6$~kPa and $p_v = 2.81$~kPa. The uncertainty of $\sigma_a$ is that of $\sigma_i$ divided by four, about $\pm 0.04$.}
  \label{tab:sigmaa}
  \end{center}
\end{table}

Taken together, the downstream displacement of the vapor is a signature of the weaker exposure of the nuclei rather than the main cause of the delay. On every surface except Rough~1, $\sigma_a$ varies linearly with the near-wall fluctuation energy, as the criterion predicts.

\section{Discussion}\label{sec:discussion}
\subsection{How the structured roughness delays inception}\label{sec:mechanism}

The results suggest the following picture. On the smooth surface, the shear layer lies directly above the nuclei migrating along the wall, so its low-pressure events cavitate them close to the wall near the throat, although the mean pressure there remains well above the vapor pressure. In the imaging, cavitation events on the smooth surface and on Rough~5 collapsed closer to the wall than on Rough~1--4, and the nuclei they left cavitated again after migrating only a short distance. On Rough~1--4, the reverse-flow layer separates the migrating nuclei from the shear layer. A plausible sequence, consistent with the imaging, is that the layer carries the nuclei upstream to its upstream end, where they turn, are lifted into the shear layer and cavitate. This would explain why the vapor forms further downstream and away from the wall (\S\ref{sec:location}).

The delay follows the fluctuation energy at the height of the nuclei, and the reverse flow is the route by which the roughness reduces it. The two vary together. From the smooth surface through Rough~5 to Rough~2--4, the near-wall reverse flow becomes more persistent, while the fluctuation energy at the nuclei and $\sigma_a$ decrease together along the criterion line (figure~\ref{fig:criterion}). These data alone cannot show which of the two controls the delay. Only the fluctuation energy, however, is linked directly to the pressure at the nuclei, through (\ref{eq:criterion2}). Rough~1 has reverse flow as strong as that on Rough~4 at $x/H = 0.15$, yet it lies above the criterion line, so the reverse flow alone does not guarantee a delay (\S\ref{sec:rough1}). Of the two events whose confluence inception requires \citep{Allan2025}, the roughness thus acts on the low-pressure event rather than on the nuclei, whose supply is similar on all surfaces (\S\ref{sec:supply}). Here the roughness weakens these events and delays inception, the opposite of the effect reported by \citet{ArndtIppen1968} for an attached boundary layer, where roughness raised the wall shear and the near-wall pressure fluctuations, and inception occurred earlier. Both results follow from the same criterion, in which inception occurs earlier when the fluctuations at the nuclei are stronger. The route to the delay also differs from that in tip-vortex flows, where roughness delays inception by weakening the vortex \citep{Asnaghi2019,Svennberg2020,Feng2026}, whereas here it lifts the shear layer away from the nuclei. Measurements over other roughness geometries and Reynolds numbers should collapse on the same relation between $\sigma_a$ and $k_{2D}/U_b^2$ at a common reference station, which would test the criterion beyond the present surfaces.

This picture extends the mechanisms of \citet{Ram_Agarwal_Katz_2020} and \citet{Russell2026}. \citet{Ram_Agarwal_Katz_2020} showed that low-momentum or separated near-wall regions shelter and recycle microbubbles, sustaining attached cavitation. \citet{Russell2026} found that microbubbles trapped in the stalled fluid ahead of a cloud cavity remain unactivated for several shedding cycles, implying that the pressure in this region is above the vapor pressure, and that the nuclei that do activate lie further from the wall or outside the stalled fluid. A few nuclei on every surface behaved in the same way here, holding their position near the throat before being drawn into the shear layer, where they cavitated. Here, the roughness-induced reverse-flow layer extends this sheltered region, so that the same near-wall fluid both sustains the supply of nuclei and delays their exposure to low pressure.

Upstream migration of nuclei occurred on all surfaces. On the smooth surface, where reverse flow is rare, it is consistent with the pressure-gradient-driven migration that \citet{Ram_Agarwal_Katz_2020} reported in a boundary layer without near-wall reverse flow (their model~II), although the present measurements do not resolve how the nuclei migrate there. On Rough~5, whose near-wall flow reverses intermittently, both routes may contribute. On Rough~1--4, where the near-wall flow is reversed on average (\S\ref{sec:flow}), the nuclei can also be carried upstream by the reverse flow, as in their separated case (model~III).

The appearance of the reverse-flow layer is consistent with the known tendency of roughness to reduce near-wall momentum and promote separation in adverse pressure gradients \citep{SongEaton2002,WuPiomelli2018}. On Rough~5 the flow reverses only intermittently and forms no mean reverse-flow layer (\S\ref{sec:reversal}), so this effect does not grow monotonically with cavity depth, the only parameter varied here.

\subsection{Rough~1: reverse flow with a marginal delay}\label{sec:rough1}

Rough~1 does not fit this picture. Its near-wall flow resembles that of Rough~2--4. It has a reverse-flow layer of similar thickness and comparable or greater strength, its shear-layer height matches theirs for $x/H \geq 0.1$, and its vapor is displaced downstream and away from the wall as on Rough~2--4 (figures~\ref{fig:vaporloc} and \ref{fig:profiles}\textit{c}). Yet its $\sigma_i$ ($3.86 \pm 0.15$) lies only marginally below the smooth value and 0.40 above the delayed group, and it lies 0.11 above the criterion line at every station and height tested (\S\ref{sec:criterion}). The offset is not explained by the uncertainty of $\sigma_i$: even at the lower bound of its $\sigma_i$, Rough~1 would lie about 0.07 above the line, nearly twice the uncertainty in $\sigma_a$. Its supply of upstream-migrating nuclei is slightly smaller than on the other surfaces (figure~\ref{fig:supply}\textit{a}), which would, if anything, delay inception, so it does not explain the offset either. Since vapor forms on Rough~1 at a similar location and under similar mean near-wall fluctuations as on Rough~2--4, the difference must lie either in what these statistics do not capture or nearer the throat. Two candidates remain. Deep low-pressure events, for example in coherent vortices, may be more frequent on Rough~1 for the same r.m.s.\ fluctuation, as \citet{RanKatz1994} found that the tail of the pressure distribution, not its r.m.s., controls inception in jets. Alternatively, the flow may differ close to the throat. On Rough~1 the near-wall flow begins to reverse closer to the throat, in 62\,\% of the samples at $x/H = 0.1$ compared with 28--45\,\% on Rough~2--4 (\S\ref{sec:flow}). At this station, however, the possible bias of the near-wall mean velocity, up to about $0.06\,U_b$ (appendix~\ref{appB}), is comparable to the reverse flow, so this difference suggests that the flow reverses earlier on Rough~1 but does not establish it. Its near-wall fluctuation energy is lower than on Rough~2--4, $k_{2D}/U_b^2 = 0.008$ at $y/H = 0.01$ against 0.012--0.040, so the mean fluctuations measured there do not explain the offset. A difference may still exist closer to the throat, at $x/H \leq 0.075$, where the present measurements are not resolved, or across the span, which the mid-plane measurements do not capture. Distinguishing between these possibilities requires the statistics of the extreme low-pressure events near the nuclei, and measurements closer to the throat and across the span.

\subsection{Rough~5: intermittent reverse flow with a smaller delay}\label{sec:reversal}

The delay is smaller at the largest roughness: on Rough~5 ($k/d = 0.72$), $\sigma_i$ rises to 3.79 and vapor forms only slightly downstream of that on the smooth surface (upstream edge at $x/H = 0.089$ versus 0.068). Rough~5 has the deepest cavities, yet its near-wall flow has no mean reverse-flow layer. It reverses only intermittently, in up to about a third of the samples, and its mean near-wall velocity is close to zero (\S\ref{sec:flow}). Why the flow over the deepest cavities stays close to separation without separating cannot be resolved from the present mid-plane measurements. Without a mean reverse-flow layer, the fluctuations at the height of the nuclei stay stronger than on Rough~2--4, though weaker than on the smooth surface, and the delay is correspondingly smaller. Rough~5 thus lies on the criterion line with the smooth surface and Rough~2--4 (figure~\ref{fig:criterion}). Its smaller delay is therefore explained by its near-wall flow alone, without an additional mechanism.

We also examined whether the surface itself contributes an additional mechanism. Gas trapped in surface crevices can serve as nuclei, with a stability that depends on the crevice geometry and on wettability \citep{Atchley1989,Borkent2009}, and bubbles growing from the etched cavities were observed only on Rough~4 and Rough~5, the surfaces with the deepest cavities (\S\ref{sec:inception_mechanisms}). These nuclei do not appear to set the inception condition. Rough~4 is the most delayed surface and Rough~5 one of the least delayed, although surface nuclei were observed on both, whereas Rough~2 and Rough~3 are delayed as much as Rough~4 without them. Rough~5 had more of these events, 14 against 6 over 20\,000 frames, but on both surfaces they amount to well under 1\,\% of the nuclei migrating upstream, they did not sustain cavitation, and Rough~5 lies on the criterion line without them. Wettability does not separate the surfaces either. All surfaces are wetting, and their contact angles agree within the drop-to-drop scatter, except possibly on Rough~4 (table~\ref{tab:roughness_surfaces}). Rough~2 and Rough~5, in particular, have nearly the same angle, but one is among the most delayed surfaces and the other among the least delayed. Surface nuclei may therefore start individual inception events, but they are not needed to explain how $\sigma_i$ varies with roughness.

\subsection{Limitations}\label{sec:limitations}

Several limitations qualify these conclusions. The velocity fields are planar and measured at the mid-plane, whereas the checkerboard cavities are three-dimensional, and $k_{2D}$ includes only in-plane fluctuations. Pressure was not measured near the wall, so the link between the lifted shear layer and weaker low-pressure events at the nuclei is inferred from $k_{2D}$ and $\sigma_i$, and the pressure-recovery estimate is one-dimensional. The supply of nuclei and the velocity and vapor fields come from separate recordings at the same operating condition. Heights in the velocity fields are resolved to about $\pm 0.002H$, and the conclusions rely only on differences of at least $0.005H$. Close to the throat ($x \leq 0.75$~mm), the reverse-flow layer is thinner than the effective resolution of the optical-flow method, and heights there are not interpreted. All flow statistics, including the reverse-flow fraction, are computed from the liquid phase only. They therefore under-sample the strongest low-pressure events where vapor is present, and rows occupied by vapor more than 70\,\% of the time are excluded. Vapor is detected by an intensity threshold. Small, fixed brightness differences between pixel rows can bias this detection in individual rows, so vapor occupancies are compared only as averages over bands of several rows.

The dissolved-gas content of the water was not measured, but it was controlled by a free surface in the storage tank, and all surfaces were tested with the same water, so it acts as a common condition rather than as a difference between surfaces. Inception was identified visually rather than from a measured event rate, but the scatter between recordings is included in the uncertainty of $\sigma_i$, $\pm 0.12$--$0.15$ (95\,\% confidence), which is about a fifth of the delay. Finally, all measurements were made at a single Reynolds number, $\Rey = 95\,160$, in one Venturi geometry and with one checkerboard pattern ($d = 384~\mu$m), varying only the depth of the cavities. Other roughness geometries may behave differently. In an attached boundary layer, for example, \citet{ArndtIppen1968} found that roughness makes inception occur earlier.

\section{Conclusion}\label{sec:conclusion}

Cavitation inception was studied in a Venturi whose diverging wall carried structured rough surfaces with roughness aspect ratios $0.03 \leq k/d \leq 0.72$, at a fixed Reynolds number of 95\,160. The main conclusions are as follows.

\begin{enumerate}
  \item The structured roughness delays inception, most strongly at intermediate roughness. Smooth walls recessed by the etched volume show no delay, so the delay requires the roughness itself.
  \item The roughness does not change the supply of nuclei. The nuclei migrating upstream along the wall are similar in number and size on all surfaces. Bubbles grown in the etched cavities are rare, and wettability differs little between surfaces, so neither explains the differences in $\sigma_i$.
  \item On Rough~1--4, the roughness creates a near-wall mean reverse-flow layer beneath a lifted shear layer, which persists in frames without vapor and weakens the fluctuations at the height of the nuclei. On Rough~5 the near-wall flow reverses only intermittently.
  \item Where the reverse-flow layer exists, vapor at inception forms further from the throat and away from the wall, but mean pressure recovery there accounts for at most about 15\,\% of the delay.
  \item On all surfaces except Rough~1, the local cavitation number at the vapor edge varies linearly with the near-wall fluctuation energy, consistent with inception triggered by pressure excursions about 6--8 times the r.m.s.\ fluctuation. The roughness therefore delays inception not by limiting the nuclei, but by weakening the fluctuations that reach them.
\end{enumerate}

The reverse-flow layer accompanies every strong delay but does not guarantee one. Rough~1 has the layer but only a marginal delay, and it lies above the criterion line. Resolving this case requires measurements of the near-wall pressure and of the three-dimensional flow within and above the cavities.

\begin{bmhead}[Acknowledgments.]
The authors thank the ONR program manager, Dr. Yin Lu (Julie) Young. This research used resources of the Advanced Photon Source, a US Department of Energy (DOE) Office of Science user facility operated for the DOE Office of Science by Argonne National Laboratory under contract DE-AC02-06CH11357. The authors are grateful to Kamel Fezzaa, Samuel Clark and the other staff of beamline 32-ID-B for their help in setting up the experiment. They also thank Ian Mitchell for the contact-angle measurements. The authors used Claude (Anthropic), accessed through the Claude web application in 2026, to assist with language editing, drafting parts of the text and writing data-analysis scripts. All work was directed, checked and approved by the authors, who take full responsibility for the content of the paper.
\end{bmhead}

\begin{bmhead}[Funding.]
This work was supported by the Office of Naval Research (ONR) under award N00014-22-1-2642.
\end{bmhead}

\begin{bmhead}[Declaration of interests.]
The authors report no conflict of interest.
\end{bmhead}

\begin{bmhead}[Data availability statement.]
The data and analysis scripts that support the findings of this study are available from the corresponding author upon reasonable request.
\end{bmhead}

\begin{appen}

\section{Image processing and track classification}\label{appA}

\subsection{Image processing}

For the tracking of nuclei, the background was subtracted from each side-view image to isolate the nuclei and vapor, which appear dark on the bright background. The images were then thresholded to give binary masks.

The masks were used for tracking in TrackMate \citep{Tinevez2017}. Candidate spots were found with its thresholding-based detector and filtered by quality and spot features before tracking. Trajectories were built with the Advanced Kalman tracker, with the initial and default search radii set from the measured frame-to-frame displacements of bubbles near the throat. Gaps of up to two frames were allowed, and tracks classified as upstream-migrating were restricted to at most one gap (appendix~\ref{app:tracking}). Track splitting and merging were enabled to capture the breakup and coalescence of bubbles. The tracks were exported as XML files and analyzed in MATLAB.

The composite images were made in ImageJ by combining the selected frames into a stack and applying the Z~Project command with the minimum-intensity option (Image~$\rightarrow$~Stacks~$\rightarrow$~Z~Project~$\rightarrow$~Min Intensity). The projection keeps, at each pixel, the darkest value over the selected frames, so that bubbles and vapor, which appear dark, are retained from every frame while the bright liquid background is unchanged.

\subsection{Track classification}\label{app:tracking}

A track is classified as upstream-migrating if it meets five criteria. It contains at least four detections, with at most one gap. Its net streamwise displacement is at most $-0.08$~mm, at least 65\,\% of its steps have $\Delta x < 0$, and at most 30\,\% have $\Delta x > 0$. No step exceeds 0.1~mm in the wall-normal direction. It starts in the downstream three quarters of the range of track starting positions, and outside a $0.5 \times 1.2$~mm region at the throat corner. None of its detections lies within a vapor region of the image. Error bars on track counts are $\pm\sqrt{N}$, one standard deviation of a Poisson count.

\section{Uncertainty of the velocity fields}\label{appB}

The accuracy of the velocity fields was estimated a posteriori with an image-matching approach based on \citet{Sciacchitano2013} and \citet{Wieneke2015}. For each image pair, the second image is warped back with the measured displacement field so that it should match the first. The normalized cross-correlation between the first image and the warped image is computed over $8 \times 8$~px windows on a 4~px grid, for integer shifts of up to $\pm 4$~px, and its peak is located to sub-pixel precision with a three-point Gaussian fit. The offset of this peak from zero shift is the residual displacement, the part of the image motion that the measured field does not capture. Only windows in the liquid, defined by the same vapor mask as the statistics, were evaluated, and windows whose peak fell on the edge of the search range were discarded. In tests with known uniform shifts imposed on liquid images, shifts of 0.5--2~px were recovered to within 0.003~px, a shift of 0.25~px was recovered as 0.17--0.19~px, and the noise floor was 0.06--0.09~px.

The residual was evaluated on 5000 image pairs per surface, spread evenly over each record, and changed by less than $0.006\,U_b$ when only half of them were used. Its r.m.s.\ value is $0.07$--$0.09\,U_b$ over the liquid field and $0.09$--$0.13\,U_b$ near the wall ($0 \leq y/H \leq 0.03$), comparable to the reverse-flow velocities. The residual is, however, largely random, its mean being typically an order of magnitude smaller than its r.m.s.\ value. The random part averages out over the record (\S\ref{sec:piv}), so the mean residual estimates the bias of the mean velocity due to image matching. Near the wall, for $x/H \geq 0.15$, it is at most $0.04\,U_b$ on Rough~1, Rough~2 and Rough~4, and at most $0.07\,U_b$ on Rough~3, whereas the reverse-flow velocities there are $0.07$--$0.16\,U_b$, at least twice as large. On the smooth surface and Rough~5, where no mean reverse-flow layer is found, it is at most $0.04\,U_b$ and $0.08\,U_b$, so a weak mean reverse flow smaller than these values cannot be excluded there. Closer to the throat ($x/H \leq 0.125$) it reaches $0.06\,U_b$, comparable to the weak reverse flow at those stations. Biases common to both images, such as vibration of the test section, are not detected by this approach. Random errors of similar size on all surfaces would raise $k_{2D}$ by the same amount everywhere, which shifts only the intercept of (\ref{eq:criterion2}), not its slope.

\end{appen}

\bibliographystyle{jfm}
\bibliography{jfm}


\end{document}